\documentclass[aps, prb, preprint, showpacs, amssymb, amsmath, superscriptaddress]{revtex4-1}
\usepackage{mathrsfs}
\usepackage{inputenc}
\usepackage{bm}
\usepackage{times}
\usepackage{graphicx}
\usepackage[colorlinks,citecolor=blue,linkcolor=blue]{hyperref}
\usepackage{array}
\usepackage{float}
\usepackage{colortbl}
\usepackage{multirow}
\usepackage{tabularx}
\begin{document}
\title{Topological semimetal phases in a family of monolayer X$_3$YZ$_6$ (X=Nb,Ta, Y=Si,Ge,Sn, Z=S,Se,Te) with abundant nodal lines and nodes}

\makeatletter
\renewcommand*{\@fnsymbol}[1]{\ensuremath{\ifcase#1\or *\or \dag\or \dag \or
    \mathsection\or \mathparagraph\or \|\or **\or \dag\dag
    \or \dag\dag \else\@ctrerr\fi}}
\makeatother
\author{Xing Wang}
\affiliation{State Key Laboratory of Metastable Materials Science and Technology $\&$ Key Laboratory for Microstructural Material Physics of Hebei Province, School of Science, Yanshan University, Qinhuangdao 066004, China}
\affiliation{College of Science, Hebei North University, Zhangjiakou 07500, China}
\author{Wenhui Wan}
\affiliation{State Key Laboratory of Metastable Materials Science and Technology $\&$ Key Laboratory for Microstructural Material Physics of Hebei Province, School of Science, Yanshan University, Qinhuangdao 066004, China}
\author{Yanfeng Ge}
\affiliation{State Key Laboratory of Metastable Materials Science and Technology $\&$ Key Laboratory for Microstructural Material Physics of Hebei Province, School of Science, Yanshan University, Qinhuangdao 066004, China}
\author{Yong Liu}
\email{yongliu@ysu.edu.cn,or ycliu@ysu.edu.cn}
\affiliation{State Key Laboratory of Metastable Materials Science and Technology $\&$ Key Laboratory for Microstructural Material Physics of Hebei Province, School of Science, Yanshan University, Qinhuangdao 066004, China}
\date{\today}


\begin{abstract}
The electronic and topological properties of single-layer X$_3$YZ$_6$ (X=Nb,Ta, Y=Si,Ge,Sn, Z=S,Se,Te) materials have been studied with the aid of first principles calculations. This kind of materials belong to topological semimetals (TMs) with abundant nodal lines and nodes. Considering their similar properties, we focus on the analysis of Ta$_3$SnTe$_6$ and Ta$_3$SiSe$_6$. The present of spin-orbit coupling (SOC) leads to the transition from type-I nodal lines to Dirac points as well as the disappear of type-II Dirac points. The three-dimensional (3D) band diagrams reproduce vividly the characteristics of nodes and nodal lines. The appearance of the flat bands in (110) edge states further confirm their nontrivial topological properties.  We also explore the relationship among different nodal lines (nodes), crystal symmetry and SOC. The type-I nodal lines are protected by M$_z$ and $\tilde{M_y}$ symmetry in the absent of SOC. Symmetry breaking leads to band splitting even in the presence of SOC. The single-layer X$_3$YZ$_6$ can be used as candidates for two-dimensional (2D) TMs and provide a platform for further study of interesting physical phenomena.
\end{abstract}

\maketitle
\section*{I. Introduction}
With the development of topological materials, topological semimetals (TSMs) have become one of the hottest topics in condensed matter physics\cite{gra1, gra2, d1,d2,d3,wl1,wl2,nl1,nl2,nl3,nl4,sm1,sm2,sm3,mp1,mp2}, originating from the character of  symmetrically or topologically protected band degeneracy near Fermi energy and potential applications. Based on the band crossing points near the Fermi energy, TSMs can be divided into Dirac semimetals (DSMs)\cite{d1,d2,d3}, Weyl semimetals (WSMs)\cite{wl1,wl2}, nodal line semimetals (NLSMs)\cite{nl1,nl2,nl3} and multiple degenerate points semimetals\cite{sm1,sm2,mp1,mp2}. DSMs have fourfold degenerate band touching points with Dirac cones surface states. WSMs possess twofold degenerate band crosspoint with Fermi-arc surface states. NLSMs are the special cases of DSMs and WSMs with continuously distributed band intersections and drumhead surface states, forming closed ring, nodal links, nodal knots or nodal chains\cite{nl4}. Na$_3$Bi, Cd$_3$As$_2$\cite{ex1,ex2} and TaAs family\cite{wl2,taas1,taas2,taas3} are the earlier experimentally discovered DSMs or WSMs materials. In addition,  NLSMs have also been verified by experiment\cite{pbtase,comnga}and host many unusual behavior, such as ultrahigh mobility\cite{1},  possible high temperature superconductivity\cite{2} and the abnormal optical response\cite{3}. Although more and more TSMs are theoretically predicted,  most of them lack experimental validation, especially the two-dimensional (2D) NLSMs, whose experimental observation conditions are more demanding. Therefore, it is urgent to propose the TSMs candidate materials for experimental implementation.\\

Recently, Nb$_3$GeTe$_6$ and Nb$_3$SiTe$_6$ have attracted a great deal of attention because of their nontrivial topological properties\cite{ta1,ta2,ta3,ta4,ta5,ta6,ta7}. They belong to van der Waals (vdW) materials with two-layer ternary telluride compounds, so the single-layer materials may obtain easily by using mechanical exfoliation method\cite{exf}. In fact, the bulk single crystals and thin sheets have been synthesized successfully in the laboratory\cite{ta5,ta8}. The bulk Nb$_3$GeTe$_6$ and Nb$_3$SiTe$_6$ are hourglass semimetals and their nontrivial topological states are guaranteed by nonsymmorphic symmetry\cite{ta1,ta5,ta7}. The present of spin-orbit coupling (SOC) make their monolayer structure become from NLSMs to DSMs\cite{ta1,ta7}. So far, there are few researches on the monolayer structures of same family or only focused these two materials. On the other hand, the analysis of single-layer topological states is not incomprehensive. Starting from the same family elements, we do more comprehensive researches of these single-layer system in order to find some interesting results.\\

In this work, we study systematically the electronic properties of single-layer X$_3$YZ$_6$ materials by using first principles calculations. They are all TSMs with rich nodes and nodal lines. X$_3$YZ$_6$ family have similar properties close to Ta$_3$SiSe$_6$ except Ta$_3$SnTe$_6$, so we take these two materials as examples. The ab initio molecular dynamics (MD) simulations and the phonon calculations suggest they are all thermally and dynamically stable. Without SOC, they have coexisting type-I nodal lines and type-II Dirac points. After considering SOC, the nodal lines transform to Dirac points and original Dirac points disappear. The local gap, Fermi surface, Fermi velocity, three-dimensional (3D) band structures and edge states further confirm their nontrivial topological properties. Finally, we prove the nodal line and some nodes are protected by crystal symmetry in the absence of SOC. These results may provide platforms to explore novel physical phenomena and topological states.

\section*{II. COMPUTATIONAL DETAILS}
The electronic properties of X$_3$YZ$_6$ is studied by using the Vienna ab initio Simulation Package (VASP)\cite{38,39}. The generalized gradient approximation (GGA)\cite{40} of Perdew-Burke-Ernzerhof (PBE)\cite{41} is treated as the exchange correlation interaction. The monolayer materials extend along the x-y plane, vacuum slab above 15 {\AA} along \textit{z}-direction is applied to exclude interactions between neighboring images. The Brillouin zone (BZ) is sampled using $10\times6\times1$ $\Gamma$-centered Monkhorst-Pack grid\cite{42}, the plane-wave cutoff energy is set to 500 eV with the energy precision of 10$^{-5}$ eV.   The atomic coordinates are relaxed until the maximum force on each atom is smaller than 0.01 eV/{\AA}. The hybrid functional HSE06\cite{43} is used to check the results and the resulting band gaps are very similar. The tight binding matrix elements are calculated by projecting the Bloch states onto maximally localized Wannier functions (MLWFs)\cite{47,48} using the VASPWANNIER90 interface. The X-d, Y-p and Z-p orbitals are used to build the MLWFs by using the Wannier90 code\cite{47,48}. Topological properties are analysed by using the iterative Green's function method as carried out in the WannierTools package\cite{49}. The phonon spectra are calculated using density functional perturbation theory (DFPT) PHONOPY code interfaced with VASP\cite{phon}.

\section*{III. RESULTS AND DISCUSSION}
\begin{figure}[htp!]
\centerline{\includegraphics[width=0.8\textwidth]{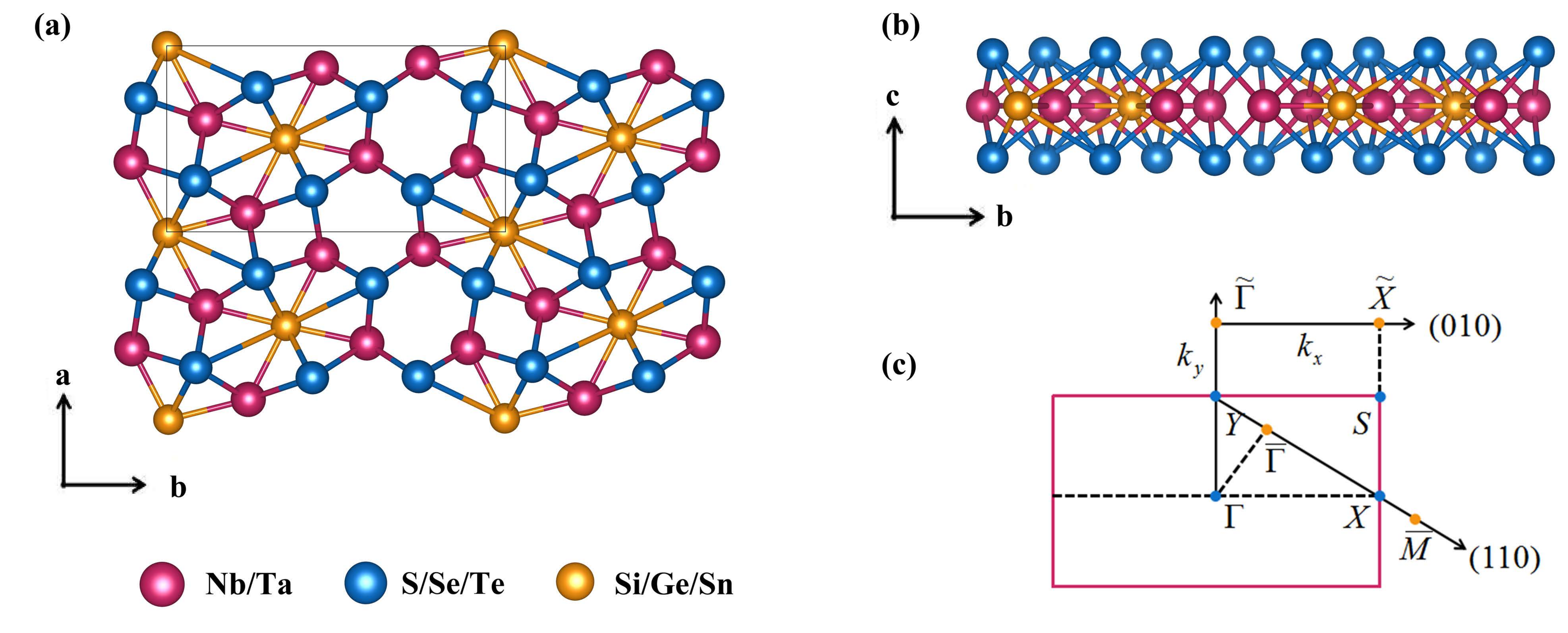}}
\caption{(Color online) The  top view (a) and  side view (b) of relaxed single-layer X$_3$YZ$_6$. The primitive cell is shown in black rectangle. (c) 2D and projected edge  first BZ with high symmetry points. }
\end{figure}

 The X$_3$YZ$_6$ (X=Nb,Ta, Y=Si,Ge,Sn, Z=S,Se,Te) monolayer has the orthorhombic structure with Pmc21 ($D_{2v}^{2}$) space group as shown in Fig. 1. The optimized lattice constants of X$_3$YZ$_6$ are nearly same to the reference values\cite{ta1} [Table I]. While fixed X element, the lattice constants have increase trend with increasing atomic radius. While fixed Y and Z elements, the lattice constants of Nb$_3$YZ$_6$ are slightly bigger than Ta$_3$YZ$_6$, which are similar to the X$_3$SiTe$_6$'s\cite{ta1}. Monolayer crystals have nonsymmorphic symmetries with a mirror operation M$_z$ ((x,y,z) $\rightarrow$ (x,y,-z)) and a glide mirror $\tilde{M_y}$ ((x, y, z) $\rightarrow$ (x+$\frac{1}{2}$, -y, z)). These symmetries provide protection for the topological NLs and Dirac points as discussed latter.

 To estimate the feasibility of preparing these monolayers, the cleavage energies are calculated [Table I]. The calculated cleavage energy for X$_3$YZ$_6$ is between 0.44 and 0.63 J/m$^2$, a little bigger than graphene (0.37 J/m$^2$)\cite{graphite} and much smaller than CaN$_2$ (1.14 J/m$^2$)\cite{can}. The graphene have exfoliated successfully in experiment,  monolayer X$_3$YZ$_6$ may expected to obtain in the same way.  Moreover, we examine thermal stability of X$_3$YZ$_6$ by performing ab initio MD simulations.  The average of the total potential energy stays the same after heating at 300 K for 10 ps [Fig. 2 (a) and supplementary material (SM) Fig. S1]. These results indicate clearly the materials remain thermal dynamically stable at room temperature. We also present the Ta$_3$SnTe$_6$ monolayer at 1000K and find the structure can still be maintained, so this monolayer has very high thermal stability and may be used to build nanoelectronic devices.  The phonon spectra of Ta$_3$SnTe$_6$ and  Ta$_3$SiSe$_6$ are plotted in Fig. 2 (b) and Fig. 6 (a) . The absence of imaginary mode in their spectra indicates dynamical stability. The single-layer X$_3$YZ$_6$ have very similar properties, so we mainly discuss Ta$_3$SnTe$_6$ and Ta$_3$SiSe$_6$ as examples.

 We have plotted the electron localization function (ELF)\cite{elf} with an isosurface of 0.78 in Fig. 2 (c). The ELF value is around 0.78 near Sn atoms and Te atoms, which means that electrons are more localized towards Sn and Te atoms. This type of electron localization indicates the ionic type of bonding between Ta and Sn (Te) atoms. To qualitatively analyze the charge transfer of Ta-Te (or Ta-Sn) bond, difference charge density map is plotted in Fig. 2 (d), the yellow/gray region represents charge accumulation/depletion, respectively. The difference pattern indicates the major charge transfer is from Ta atom to Te (or Sn) atom, which is consistent with the characteristics of an ionic bond. The similar results for Ta$_3$SiSe$_6$ and Ta$_3$SiS$_6$ are presented in Fig. S2.

\begin{table*}[!htbp]
\centering
\caption{The lattice constants \textit{a}, \textit{b} and cleavage energy E$_c$ of single-layer X$_3$YZ$_6$.}
\begin{tabular}{c c c c | c c c c} 
\hline
\hline
Structure&$\textit{a}$({\AA})&$\textit{b}$({\AA})&E$_c$({J/m$^2$})&Structure&$\textit{a}$({\AA})&$\textit{b}$({\AA})&E$_c$({J/m$^2$})\\
\hline
Ta$_3$SiS$_6$&5.825&10.851&0.53&Nb$_3$SiS$_6$&5.835&10.961&0.46\\
\hline
Ta$_3$SiSe$_6$&6.050&11.105&0.44&Nb$_3$SiSe$_6$&6.069&11.210&0.44\\
\hline
Ta$_3$SiTe$_6$&6.402&11.559&0.44&Nb$_3$SiTe$_6$&6.410&11.634&0.63\\
\hline
Ta$_3$GeTe$_6$&6.522&11.637&0.49&Nb$_3$GeTe$_6$&6.536&11.706&0.50\\
\hline
Ta$_3$SnTe$_6$&6.734&11.806&0.42&Nb$_3$SnTe$_6$&6.745&11.880&0.50\\
\hline
\hline
\end{tabular}
\end{table*}

\begin{figure}[htp!]
\centerline{\includegraphics[width=0.8\textwidth]{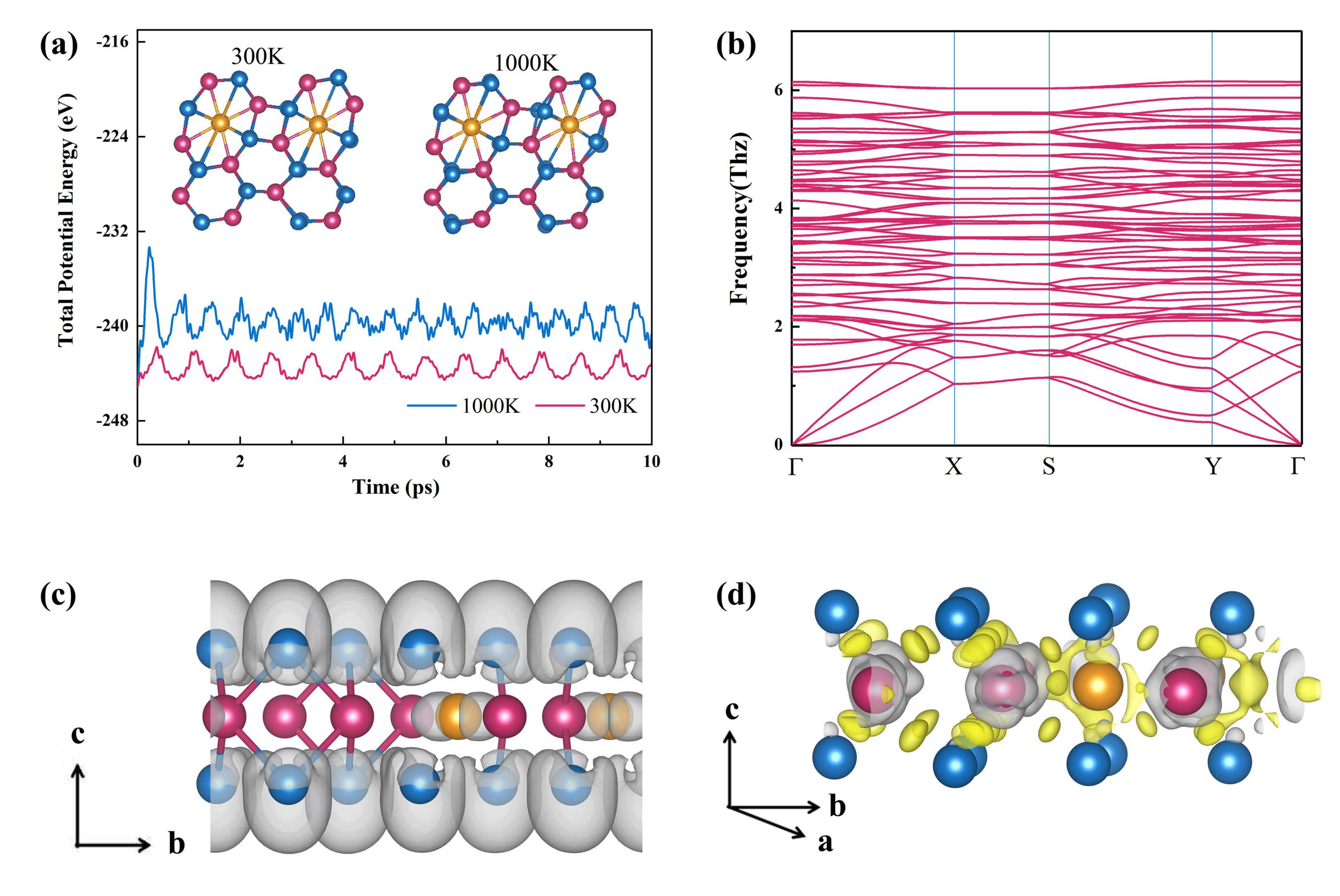}}
\caption{(Color online) MD simulations (a), phonon spectra (b), ELF (c) and difference charge density (d) of single-layer Ta$_3$SnTe$_6$. (a) Snapshots of atomic configurations at the end of MD simulations and total potential energy fluctuations observed at 300K and 1000K, respectively. (c) Isosurface corresponding to ELF value of 0.78. (d) The yellow (gray) isosurface plots correspond to the charge density accumulation (depletion). Isosurface corresponding to difference charge density of $\pm$ 0.01 eV/${\AA}^3$.}
\label{fig:fig2}
\end{figure}

\begin{figure}[htp!]
\centerline{\includegraphics[width=1.0\textwidth]{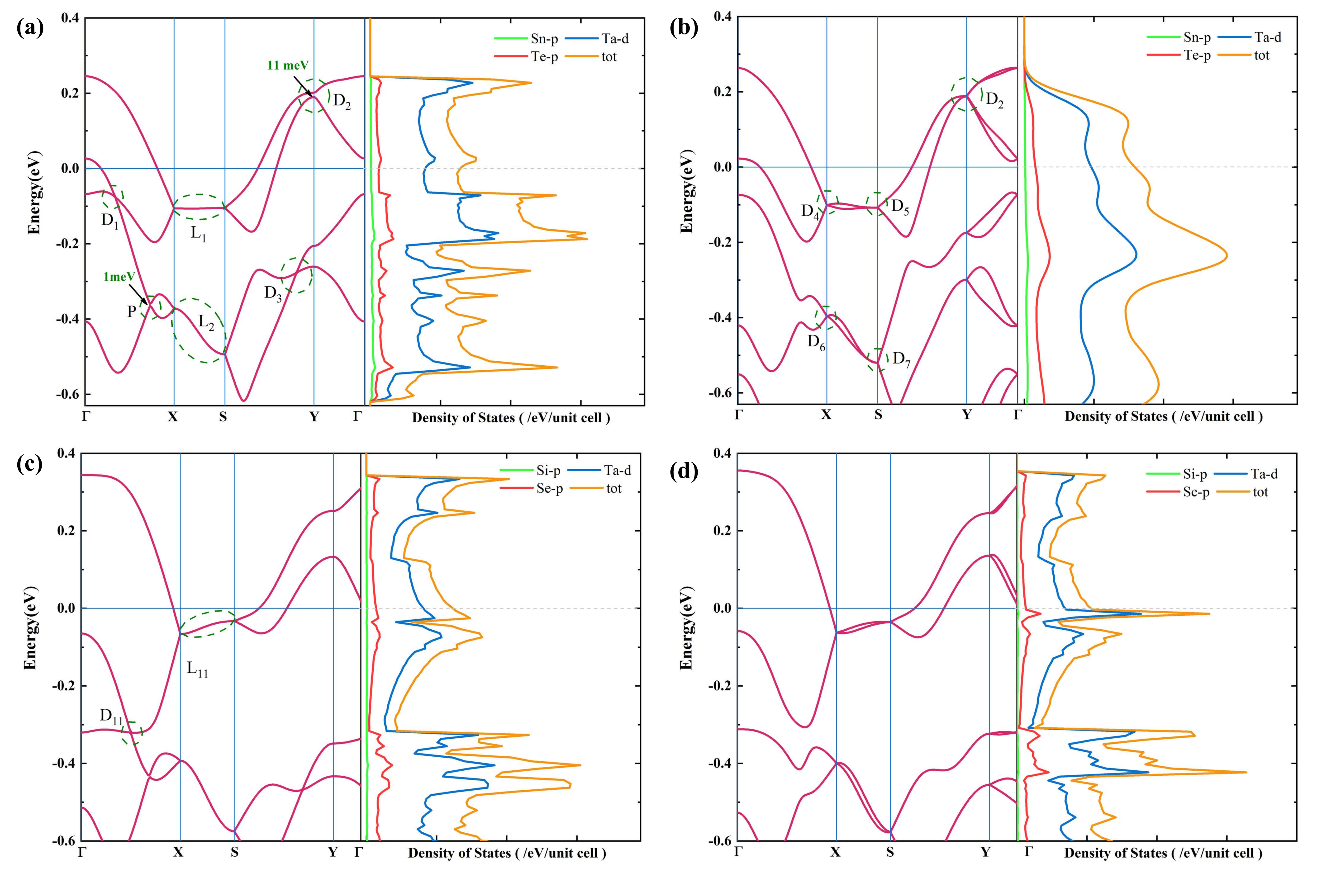}}
\caption{(Color online) Band structures and  projected density of states (PDOS) of single-layer Ta$_3$SnTe$_6$ (a, b) and Ta$_3$SiSe$_6$
(c, d), respectively. (a, c) without considering the SOC. (b, d) with considering the SOC. The Fermi levels are set to zero.}
\label{fig:fig3}
\end{figure}

 Here, we have calculated the band structures and projected density of states (PDOS). In the absence of SOC, the Ta$_3$SnTe$_6$ displays a semimetal, the valence and conduction bands meet in the vicinity of the Fermi level and  show a nodal line L$_1$, as plotted in Fig. 3 (a). The energy of the nodal line L$_1$ varies with 2 meV (flat) and it is not coexisting with other trivial bands (clean), making it an ideal NLSMs candidate material\cite{shm}. Moreover, there is a small band gap of 11 meV at high symmetry points Y labelled D$_2$. The red line and point in the picture of the local gap near the Fermi level confirm the existence of L$_1$ and D$_2$, shown in Fig. 4 (a). Nodes and nodal lines are not confined to Fermi energy, but exist in a certain range near Fermi energy, which also appears in other topological semimetals\cite{b1,b2}. Additional band crossing are observed below the Fermi level in energy range of 0.5 eV, labelled two Dirac points D$_1$, D$_3$ and a nodal line L$_2$\cite{b1,b2,b3}. The obtained Fermi surfaces on the (001) cleaved surface are plotted in Fig. 4 (b), which is consistent with the previous analysis. In the middle, there are two long wavy lines around nodal line L$_1$ and unsealed oval chain structure surrounding the D$_2$. The band structures of the Ta$_3$SiSe$_6$ indicate it is also a semimetal in Fig. 3 (c), the nodal lines and nodes are similar to the case of Ta$_3$SnTe$_6$. The middle red line show the existing of nodal line L$_{11}$ in Fig. 6 (b), but red point disappears because the band gap at point Y is larger, which leads to closed ellipse Fermi surface structure around it in Fig. 6 (c). The band structure diagrams of the remaining structures of X$_3$YZ$_6$ are shown in the Fig. S3, their electronic properties are more closer to case of Ta$_3$SiSe$_6$. To illustrate the unique electronic property, we have plotted the PDOS [Fig. 3 (a, c)], which agree well with the band structure distribution. The low-energy states around Fermi level are mainly from the Ta d orbitals, the contribution of Si (or Sn) and Se (or Te) p orbitals are very small.

To reveal the energy dispersion of the nodes and nodal lines, the 3D band structures of Ta$_3$SnTe$_6$ and Ta$_3$SiSe$_6$ are plotted in Fig. 5 (a, c, d) and Fig. 6 (d), respectively. The valence band maximum (VBM) and conduction band minimum (CBM) meet in the vicinity of Fermi level with opposite slopes, which shows  type-I\cite{type1,type2} nodal lines L$_1$ and L$_{11}$. The two bands form nodal line L$_2$ with opposite slopes, so L$_2$ is also a type-I nodal line. Although they are all type-I nodal lines, the shapes of the nodal lines are different. They are straight (L$_1$), parabolic (L$_2$) and curve (L$_{11}$), respectively. The D$_1$, D$_3$ and D$_{11}$ are type-II Dirac points, because the two bands are tilted in the same slope sign. The Berry phases around nodal line L$_1$ and Dirac cone D$_1$ are $\pi$, further supporting the existence of the non-zero Berry phase and the existence of Dirac particles in Ta$_3$SnTe$_6$. The nodes and nodal lines near the Fermi surface has extremely high mobility, the Fermi velocity of D$_1$ is anisotropic and about $2.1\times 10^5$ m/s in (010) surface, compared to HgTe quantum well ($5.5\times 10^5$ m/s)\cite{vf}.

The edge states are another manifestation of the nontrivial topological properties. The edge states of (010) and (110) surface for Ta$_3$SnTe$_6$ are showing in Fig. 4 (c, d).  The corresponding positions of Dirac points and nodal line are found in the edge states. As an interesting feature of NLSMs is the appearance of flat band in the edge states\cite{flat}, we also find a flat band labelled $L_{1}^{'}$ with its band width being about 14 meV on the (110) surface edge state. Since the edges are electrically neutral, flat bands may be half-filled, providing an interesting platform for strongly correlated physics\cite{flat}. The edge states of Ta$_3$SiSe$_6$ are similar to the Ta$_3$SnTe$_6$'s, we also find the position of nodal line L$_{11}$ and node D$_{11}$ in the edge state, shown in Fig. 6 (e, f). The flat bands confirm the topological nontriviality of Ta$_3$SnTe$_6$ and Ta$_3$SiSe$_6$.

\begin{figure}[htp!]
\centerline{\includegraphics[width=1.0\textwidth]{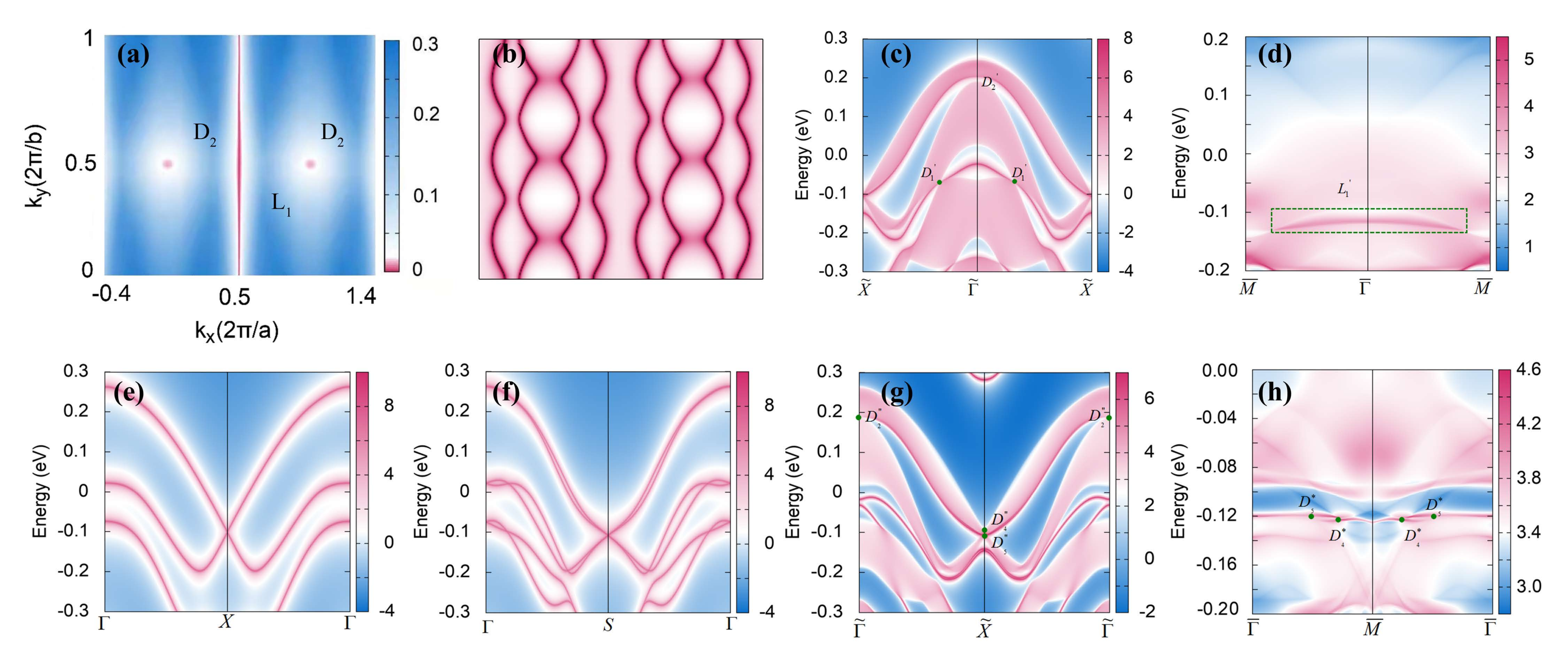}}
\caption{(Color online) The band gap (a), Fermi surface (b), edge state(c, d, g, h), the zoom-in of the band structure (e, f) for single-layer Ta$_3$SnTe$_6$. (a) The local gap near the Fermi level. (b) The calculated Fermi surface on the k$_z$=0. The edge states on the (c, g) (010) and (d, h) (110) surface, (c, d) without and (g, h) with considering SOC, respectively. The zoom-in of the band structure around the X point (e) and S point (f) with considering SOC, respectively. The Fermi level is set to zero.}
\label{fig:fig4}
\end{figure}

\begin{figure}[htp!]
\centerline{\includegraphics[width=0.8\textwidth]{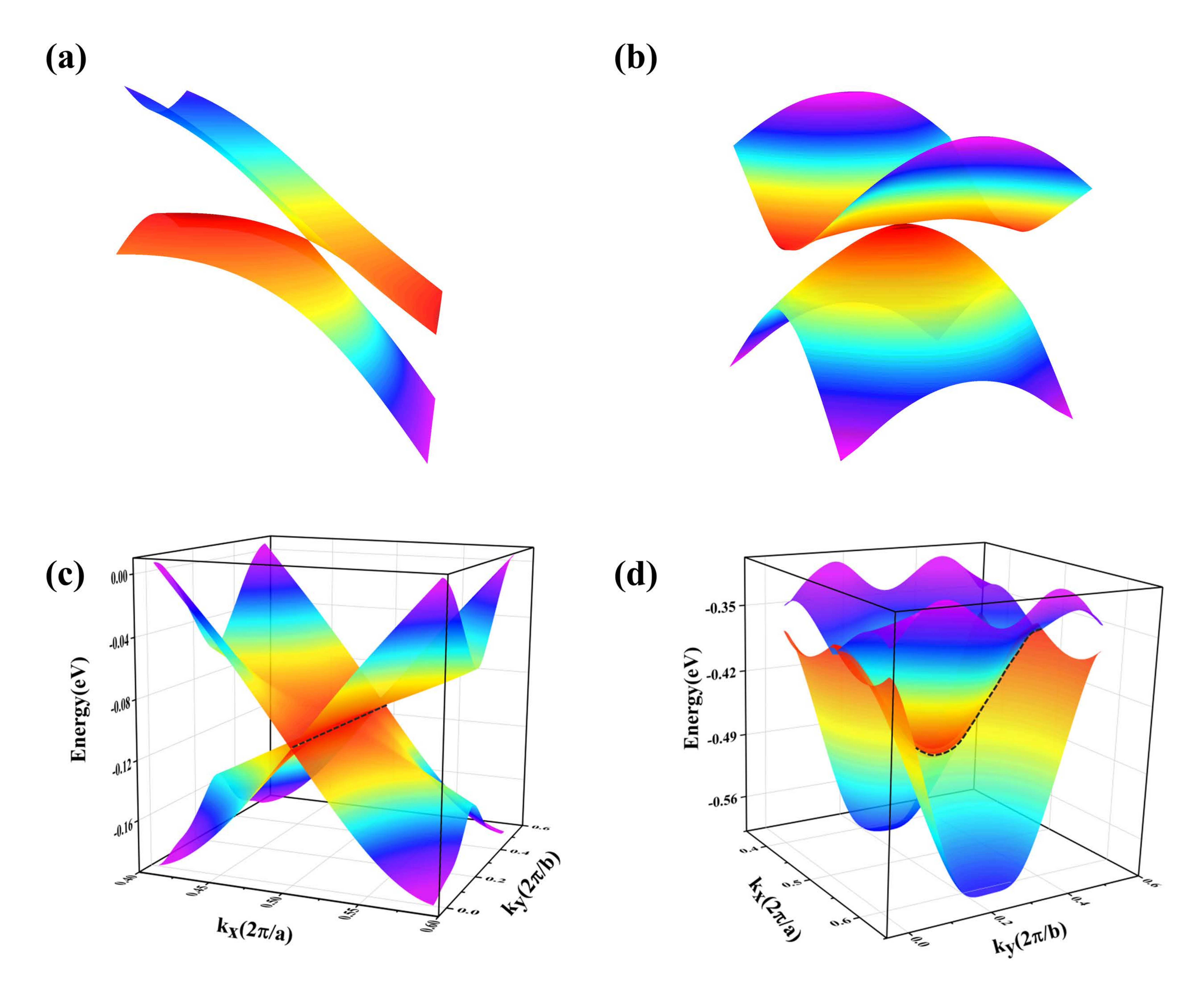}}
\caption{(Color online) 3D band structures for single-layer Ta$_3$SnTe$_6$.  3D band structures near the Dirac points D$_1$ (a) and D$_2$ (b), nodal lines L$_1$ (c) and L$_2$ (d), respectively.  Nodal lines are denoted by black dotted lines. The different colors indicate the energy values deviating from the Dirac point or nodal line.  (a, c, d) without and (b) with considering SOC, respectively. The Fermi level is set to zero.}
\label{fig:fig5}
\end{figure}
Then we discuss the stability of Dirac points and nodal lines under different cases. In the present of SOC, the X$_3$YZ$_6$ materials still display semimetal phases, the valence and conduction bands meet in the vicinity of the Fermi level [Fig. 3 (b, d) and Fig. S3]. For Ta$_3$SnTe$_6$, the nodal line L$_1$ is gapped with a small band gap about 14 meV and form two Dirac points D$_4$ and D$_5$. The case of L$_2$  is similar to L$_1$, the nodal line becomes two Dirac points D$_6$ and D$_7$.  The Fermi velocities are about $2.3\times 10^4$ m/s for D$_4$ and $2.1\times 10^3$ m/s for D$_5$ in X-S direction, which are close to case in Ta$_3$SiTe$_6$\cite{ta1}. The band intersection P disappears, the band gap of D$_2$ becomes small and forms a Dirac point. The 3D band structure around D$_2$ shows the anisotropic Dirac cone [Fig. 5 (b)].  The Fermi velocity of D$_2$ is about $8.8\times 10^3$ m/s in (010) surface. The zoom-in of the band structure around the X point and S point are plotted in Fig. 4 (e, f), the Dirac points become more obvious. The calculated edge states of the (010) and (110) surface are showing in Fig. 4 (g, h), which are another manifestation of the nontrivial topology of the Ta$_3$SnTe$_6$. One can see easily the emergence of topological nontrivial edge states and Dirac points. The changes of nodal lines for Ta$_3$SiSe$_6$ is similar to Ta$_3$SnTe$_6$'s, but the variation of band gap at the point Y is a little different. For Ta$_3$SnTe$_6$, the band gap at Y point becomes small and forms a Dirac point. But the band gap barely changes for Ta$_3$SiSe$_6$, there is no Dirac point. In fact, the X$_3$YZ$_6$ materials except Ta$_3$SnTe$_6$ have band gaps at point Y, only Ta$_3$SnTe$_6$ have an extra Dirac point D$_2$.

\begin{figure}[htp!]
\centerline{\includegraphics[width=1.0\textwidth]{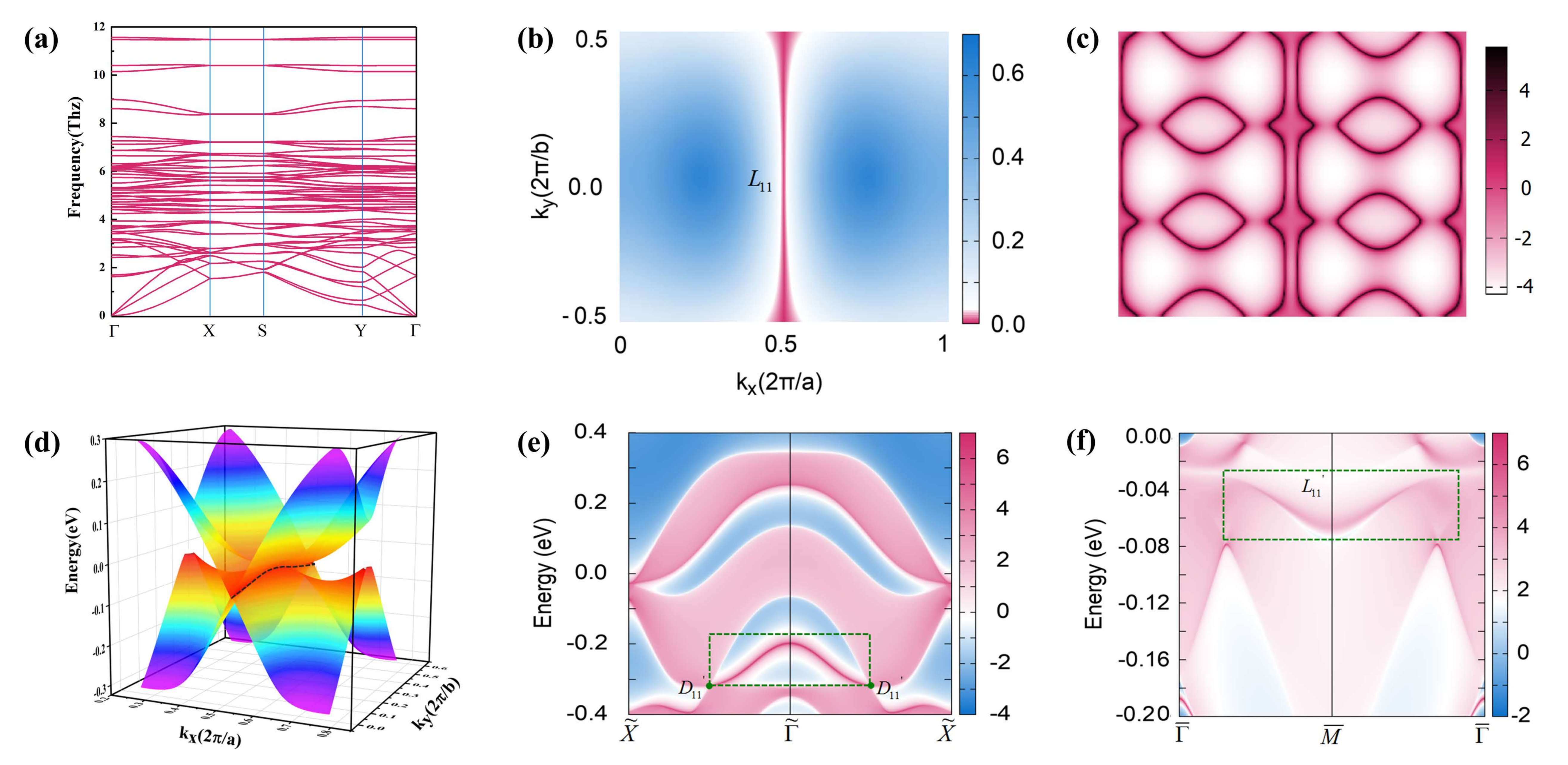}}
\caption{(Color online)  Phonon spectra(a), band gap (b), Fermi surface (c), 3D band structures (d) and edge state(e, f) for single-layer Ta$_3$SiSe$_6$ in the absence of the SOC.  (b) The local gap near the Fermi level. (c) The calculated Fermi surface on the k$_z$=0. (e) 3D band structures near the nodal lines L$_{11}$  to show its nodal line features.   Nodal line is denoted by black dotted line. The different colors indicate the energy values deviating from the nodal line. The edge states on the (e) (010) and (f) (110) surface.  The Fermi level is set to zero.}
\label{fig:fig6}
\end{figure}

To prove nodes and nodal lines are protected by symmetry, several tests are performed for Ta$_3$SnTe$_6$. First, we artificially shift the positions of the Ta atoms along the c axis to break the mirror symmetry M$_z$ and maintain the $\tilde{M_y}$. The node D$_1$, nodal lines L$_1$ and L$_2$ are all fully gapped, while node D$_3$ still exists, shown in Fig. 7 (c). These mean D$_1$, L$_1$ and L$_2$ are protected by M$_z$ symmetry in the absence of SOC. In the present SOC, the band of nodal line splits further to form two double degenerate points [Fig. 7 (e)].  Second, we shift the positions of the Ta atoms along the a axis to break the mirror symmetry $\tilde{M_y}$ and maintain the M$_z$ symmetry. The nodes D$_1$, D$_3$, nodal lines L$_1$ and L$_2$ are all fully gapped in the absence of SOC, plotted in Fig. 7 (d), which means D$_1$, L$_1$ and L$_2$ are protected by $\tilde{M_y}$ symmetry, while D$_3$ and D$_2$ are affected by $\tilde{M_y}$. With considering SOC, the band splits further [Fig. 7 (f)]. To further clarify the relationship between D$_2$ (or D$_3$) and symmetry, we perform a third test. We use 10\% uniaxial tensile strain along the a axis and b axis direction for single-layer Ta$_3$SnTe$_6$, but the crystal symmetry stays the same. The band structures are plotted in Fig. 7 (g) (h), we find the gaps of D$_1$, L$_1$ and L$_2$ are not change, but the gap of D$_2$ is changed and node D$_3$ disappears under strain along the a axis, or the gap of D$_2$ becomes zero and position of D$_3$ has changed under strain along the b axis. These results show D$_1$, L$_1$ and L$_2$ are protected by M$_z$ and $\tilde{M_y}$ symmetry, D$_2$ and D$_3$ are affected by the two symmetries.

\begin{figure}[htp!]
\centerline{\includegraphics[width=1.0\textwidth]{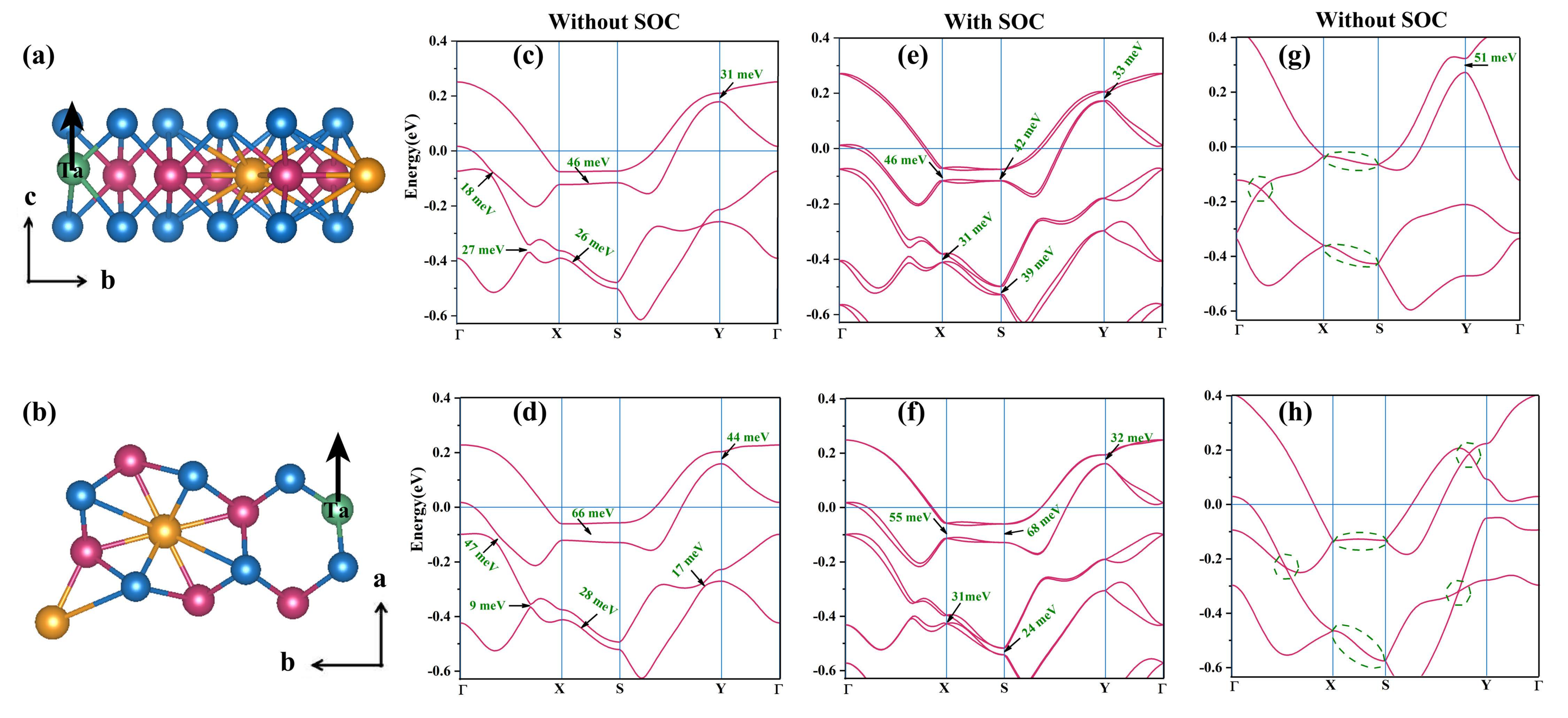}}
\caption{(Color online) Side view (a) and top view (b) of the crystal structure for single-layer Ta$_3$SnTe$_6$. Black arrows indicate the shift (0.01 {\AA}) of the Ta atoms along the c axis and a axis, respectively.  The band structures (c, d) without and (e, f) with considering SOC after moving, (c, e) and (d, f) for the configuration in (a) and (b), respectively. The band structures under 10\% uniaxial tensile strain along the a axis [(g)] and b axis [(h)] direction, respectively. The Fermi level is set to zero.}
\label{fig:fig7}
\end{figure}

To test the effect of electron correlation on band structure, we perform GGA + U calculations with considering hubbard U correction of transition metal elements. Compared with GGA, there is little change in the results of single-layer Ta$_3$SnTe$_6$ and Ta$_3$SiSe$_6$, shown in Fig. S4 and S5. In addition, we use the hybrid functional HSE06 to check the electronic structure, similar nodes and nodal lines are found for single-layer Ta$_3$SnTe$_6$ and Ta$_3$SiSe$_6$ without considering the SOC in Fig. S6. We also have analyzed the electronic structure of the two-layer Ta$_3$SnTe$_6$ and found the nodes and nodal lines will still exist, as shown in Fig. S7.

\section*{IV. Conclusion}
In summary, we find the monolayer X$_3$YZ$_6$ have similar properties and they are all TSMs materials based on the first principles calculations. The MD simulations indicate they are all thermal-dynamically stable. Taken Ta$_3$SnTe$_6$ and Ta$_3$SiSe$_6$ as examples, they have coexisting type-I nodal lines and type-II Dirac points in the absent of SOC.  With the presence of SOC, the nodal lines become Dirac points and original Dirac points disappear. The characteristics of nodal lines and nodes can be clearly seen by using 3D band diagrams. Their nontrivial topological properties are confirmed by (010) and (110) surface edge states. The relationship among different nodal lines (nodes), crystal symmetry and SOC has been discussed. D$_1$, L$_1$ and L$_2$ are protected by M$_z$ and $\tilde{M_y}$ symmetry, D$_2$ and D$_3$ are affected by these two symmetries for Ta$_3$SnTe$_6$. The SOC cause the band to split further. X$_3$YZ$_6$ materials can be ideal candidates for the nodal line and node TSMs. These results can be used as a supplement to the research content of Ta$_3$SiTe$_6$ and provide theoretical basis for subsequent experimental research.

\begin{acknowledgments}
This work was supported by National Natural Science Foundation of China (No. 11904312 and 11904313), the Project of Department of Education of Hebei Province, China(No. BJ2020015), and the Natural Science Foundation of Hebei Province (No. A2019203507 and A2020203027). The authors thank the High Performance Computing Center of Yanshan University.
\end{acknowledgments}

\end{document}